# Nonreciprocal Phased-Array Antennas

J. W. Zang[1,2,3], A. Alvarez-Melcon[1,4], J. S. Gomez-Diaz[1]*

[1]Department of Electrical and Computer Engineering, University of California Davis
One Shields Avenue, Kemper Hall 2039, Davis, CA 95616, USA.
[2]China Academy of Information and Communications Technology, Beijing 100191, China
[3]School of Information and Electronics, Beijing Institute of Technology, Beijing 100081, China
[4]Universidad Politécnica de Cartagena, 30202 Cartagena, Spain
*E-mail: jsgomez@ucdavis.edu

*A phased-array antenna is a device that generates radiation patterns whose shape and direction can be electronically controlled by tailoring the amplitude and phase of the signals that feed each element of the array. These devices provide identical responses in transmission and reception due to the constrains imposed by time-reversal symmetry. Here, we introduce the concept of nonreciprocal phased-array antennas and we demonstrate that they can exhibit drastically different radiation patterns when operated in transmission or in reception. The building block of the array consists of a time-modulated resonant antenna element that provides very efficient frequency conversion between only two frequencies: one associated to waves propagating in free-space and the other related to guided signals. Controlling the tunable nonreciprocal phase response of these elements with the phase of low-frequency modulation signals permits to independently tailor the transmission and reception radiation patterns of the entire array. Measured results at microwaves confirm isolation levels over 40 dB at desired directions in space with an overall loss below 4 dB. We believe that this concept can be extended across the electromagnetic spectrum provided adequate tuning elements are available, with important implications in communication, sensing, and radar systems, as well as in thermal management and energy harvesting.*

## I. Introduction

Phased-array antennas consist of multiple antennas appropriately arranged in space to provide tailored and highly directive radiation patterns that can be electronically controlled without the need of mechanical rotation [1-5]. They are ubiquitous in modern technology from radiofrequencies (RF) to optical frequencies and find wide applications in military radar systems and tracking platforms, civilian automotive radars [6, 7], light-detection-and-ranging (LIDAR) devices [8, 9], satellite, wireless, and optical communications [10-13], radio astronomy [14, 15], imaging [16], and remote and biological sensing [17-20], among many others. The first phased array antenna was demonstrated in 1909 by employing a three-element switchable configuration to enhance the transmission of radio waves in one direction [21]. Although there has been continuous progress in phased-array antennas during the last decades, their basic operation principle has remained essentially unchanged since its invention: the amplitude and phase excitation of each antenna element are individually tailored in such a way that the radiated waves interfere constructively in desired directions and destructively in undesired ones. The advantages of phased-array antennas over single radiating elements include significantly higher transmission gain, reception sensitivity, and power handling, as well as the ability to synthesize a large variety of radiation patterns. Additionally, ultra-rapid beam scanning and shaping can be realized by electrically manipulating the excitation of the antenna elements, usually through tunable feeding networks composed of digitally-controlled phased shifters [22]. Recently, smart antennas [23-26] have merged sophisticated processing algorithms [27] with antenna arrays to enable real-time functionalities, crucial in emerging 5G and optical communication systems. To this purpose, the amplitude and phases of the signals that feed each element of the antenna array are continuously updated as a function of the received waves. Application examples include finding the direction of arrival of unknown signals [28-30], adaptive beamforming [31], and multiple target tracking [32].

Phased-array antennas exhibit identical radiation patterns in transmission and reception due to the restrictions imposed by time-reversal symmetries [33]. Merging nonreciprocal responses with the flexibility provided by smart antennas would permit, for the first time to our knowledge, to dynamically and independently control the transmission and reception properties of the array at the same operation frequency, opening exciting venues in communication and sensing systems and also in related areas of thermal management. Such antenna would be able to efficiently handle unwanted interferences and jamming signals that might otherwise block the device; mitigate cross-talking and mutual-coupling effects that often arise in electromagnetically crowded environments, such as in the roof of buildings, ships, aircrafts, or integrated chips; enhance the channel diversity in multiple-input multiple-output (MIMO) [2] radio links; and add new knobs to boost the dynamic performance of radars, sensors, and wireless networks across the electromagnetic spectrum. Unfortunately, there are not tunable and nonreciprocal radiating elements that can serve as a building block for smart antennas systems. Early attempts to develop this type of antennas employed ferrites to break reciprocity [34-36], leading to devices that exhibited limited efficiency and whose tunable responses required the presence of bulky and lossy magnets that are not compatible with integrated circuits. Another possibility relies on using gyrators or non-reciprocal phase-shifters in the network that feed the elements of an antenna array, thus imparting different phases to the waves that are transmitted or received [37-38]. One of the major challenges of using nonreciprocal phase shifters, which usually rely on magneto-optical effects [39-42] or on active elements [43-45], is that the phase difference that they impart to waves that propagate in forward and backward directions is usually fixed and cannot be easily controlled. As a result, these components cannot be applied to realize antennas with intendent transmission and reception radiation patterns. In any case, this approach to implement non-reciprocal feeding networks in phase-array antennas has yet to be explored in practice. In a related context, the field of active integrated antennas attracted significant attention in the early years of the XXI century. There, radiating elements were combined with active and digital circuits to enable functionalities such as duplexing, mixing, amplification, and even signal processing [46-47]. More recently, magnetless spatiotemporal modulation techniques [48-50] have been applied to realize nonreciprocal leaky-wave antennas by exploiting space-time transitions between guided and leaky-wave modes [51-53]. Even though isolation levels up to 15 dB have been reported between transmission and reception in specific directions close to endfire [51], leaky-wave antennas suffer from important challenges in terms of size, complexity, efficiency, and dispersive beam scanning behavior that not only limit their use in smart antenna configurations but also in most practical scenarios [54, 55]. Finally, different types of time-modulated metasurfaces have recently been put forward to manipulate the refraction and transmission properties of beams propagating in free-space. For instance, they have been demonstrated to behave as serrodyne frequency translators employing a sawtooh waveform as modulation signal [56]. In addition, space-time coding metasurfaces have recently enabled simultaneous control of electromagnetic waves in both spatial direction and harmonic power distribution [57]. To this purpose, the amplitude or phase of the reflection coefficient associated to each unit-cell is controlled in such a way that implements a digital "0" or "1" and then an appropriate temporal coding is applied to all elements of the metasurface. Using advanced coding technique allows to further extend the range of available functionalities, including an almost independent of control of the amplitude and phase of the generated nonlinear harmonics [58] and novel architectures for wireless communication systems [59]. Similar responses have been obtained using time-modulating Huygens metasurfaces and independently tailoring in time and space the magnetic and electric dipoles that compose each unit-cell of the structure [60]. Nonreciprocal beam scanning for fixed directions in space has been theoretically investigated by inducing space-time photonic transitions in spatiotemporally modulated surfaces [61] and a more general form of the classical Snell's relation not bounded by Lorentz reciprocity was also introduced [62]. Nonreciprocal wavefront control was recently demonstrated using time-gradient modulated metasurfaces by imposing drastically different phase gradients during up and down frequency conversion process [63]. This approach permits to implement functionalities such as beam steering and focusing while providing angle-insensitive non-reciprocal responses unable to shape any beam. In all cases, time-modulated metasurfaces relate incoming and refracted/transmitted waves propagating in free-space that are not converted into guided signals. Therefore, despite recent advances in the nascent field of magnetless nonreciprocity [64-70], state of the art antennas in communication, radar, and sensing systems are unable to break and tailor reciprocity at will to enhance their performance.

In this paper, we introduce the concept of nonreciprocal phased-array antennas, which significantly extends the functionalities of smart antennas by enabling an independent and dynamic control of transmission and reception radiation patterns at the same operation frequency. To this purpose, we relate states associated to spatial and guided waves in time-modulated antennas using photonic transitions and exploit the photonic Aharonov-Bohm effect [71-73] to impart controllable nonreciprocal phases to waves that are either transmitted or received. It should be noted that each time-modulated antenna provides an extra gauge degree of freedom to the photons involved in the transitions that is associated to the arbitrary choice of time origin of the modulating signal. As a result, the nonreciprocal phases imposed on waves radiated or received by an isolated antenna depend on the choice of gauge and thus are not directly observable, in clear analogy with the electronic Aharonov-Bohm effect [74-75]. Instead of using a single antenna element, we propose here to employ an antenna array that relies on controlling the phase differences of the radiated/received waves to create constructive/destructive interferences along desired directions in space. Such phase differences are gauge invariant and thus can be detected in practice, which permits to effectively impart nonreciprocal phases to the waves transmitted and received by each antenna element.

The fundamental building block of our proposed platform is a time-modulated resonant antenna that is simultaneously excited from two ports. By appropriately imposing even and odd symmetries at nonlinear harmonics frequencies through a feedback mechanism, we show that it is possible to enforce very efficient frequency conversion between only two frequencies associated to signals guided in the structure and to waves propagating in free-space. This approach permits to implement efficient time-modulated resonant antennas in which the mixer is part of the device and takes advantage of its resonant behavior to implement photonic transitions across the electromagnetic spectrum, including the realm of infrared and optics, without relying on complex digital circuits [46-47]. The phase response of the resulting antenna element when operated in transmission or reception is controlled in a nonreciprocal manner through the phase of the low-frequency modulating signal. Nonreciprocity in the phase arises due to the photonic Aharonov-Bohm effect [71-73], in which reverting the direction of the photonic transition – i.e., from transmission to reception– changes the sign of the induced phase; and can also be understood in terms of nonlinear phase conjugation, a technique usually employed in the design of mixers [76]. For the sake of demonstration, we implement this configuration at microwaves using a simple patch antenna loaded with two varactors. Measured results confirm the nonreciprocal phase response of the element together with an excellent efficiency and an overall loss below 3 dB. Remarkable, the amplitude of all unwanted nonlinear harmonics is at least 30 dB smaller than the signals of interest. Based on this efficient nonreciprocal antenna, we demonstrate a two-element phased-array antenna able to provide over 40 dB of isolation between transmission and reception in the direction perpendicular to the structure (broadside) while exhibiting an overall loss of just 4 dB. Importantly, the time-origin of the signals that modulate the patches has been synchronized to impose the same gauge degree of freedom to all photonic transitions. As a result, the phase difference of the signals transmitted and received by each antenna element become observable and gauge independent. By simply manipulating the phases of the modulating signals, we show that it is possible to favor the transmission or reception of energy at desired directions, obtain common reciprocal radiation patterns, and implement beam scanning functionalities. Our measured results fully confirm the fundamental principles of nonreciprocal phased-array antennas. Even more sophisticated functionalities can be obtained by increasing the number of radiating elements and gathering them in two-dimensional arrangements. We emphasize that the proposed nonreciprocal antenna concept can be implemented with different technologies at any frequency band provided that adequate reconfigurable materials or components are available. We believe that this concept can also easily be extended to other fields such as thermodynamics and energy harvesting.

## II. Operation principle of nonreciprocal phased-array antennas

Consider a resonant and nonlinear antenna that is time-modulated with a signal with low frequency $f_m$ and phase $\varphi_m$. The nonlinear process occurring in the antenna generates nonlinear harmonics at frequencies $f_0 + nf_m$ (with $n \in \mathbb{Z}$). Tailoring the antenna resonant response and exploiting symmetry constrains, as described in the following, it is possible to achieve very efficient frequency conversion between only two frequencies: one associated to waves propagating in free-space and other related to the signals within the antenna feeding network. It should be stressed that this nonlinear frequency conversion process is not reciprocal, neither in phase nor in amplitude. The operation principle of the resulting time-modulated antenna, assuming frequency conversion with the first odd nonlinear harmonics ($n = \pm 1$), is as follows. In transmission (Fig. 1a, top), the antenna up-converts the excitation signals oscillating at $f_0$ to $f_0 + f_m$ ($n = +1$) and radiates them towards free-space with a phase proportional to $+\varphi_m$. In reception (Fig. 1a, bottom) the antenna receives incoming waves oscillating at $f_0 + f_m$ and down-converts them to $f_0$ ($n = -1$) with a phase proportional to $-\varphi_m$. Strong nonreciprocity in the phase appears during the transmission/reception of waves, associated to phase conjugation during up/down frequency conversion processes [76] and to the photonic Aharonov-Bohm effect [71-73]. Note that the conversion efficiency of these processes is very similar when the modulation frequency $f_m$ is significantly smaller than the operation frequency (i.e., $f_m \ll f_0$), as detailed in Appendix B.

Using time-modulated antennas as radiating elements, nonreciprocal phased-arrays with drastically different radiation patterns in transmission and reception can be constructed. Fig. 1b shows the diagram of a linear array configuration operating in transmission. The device consists of a feeding network for the signal oscillating at $f_0$, a second feeding network that incorporates phase-shifters for the low-frequency modulation signal $f_m$, and identical nonlinear antenna elements. Applying the well-known array factor approach [1-2], the electric field $\mathbf{E}_t$ radiated by the array at $f_0 + f_m$ can be approximately computed as

$$\mathbf{E}_t(\theta,\phi) = \mathbf{E}_{ant}(\theta,\phi) \sum_{p=1}^{P} w_p e^{j(\psi_p + \varphi_{mp})}, \tag{1}$$

where $\mathbf{E}_{ant}(\theta,\phi)$ denotes the radiation pattern of the individual antenna, being $\theta$ and $\phi$ the elevation and azimuth angles in spherical coordinates, respectively; $P$ is the total number of antennas in the array; $w_p$ and $\psi_p$ are the amplitude and phase of the signal $f_0$ that feed an element '$p$', and $\varphi_{mp}$ is the phase of the signal oscillating at $f_m$ that modulates the '$p$' antenna. This approach can easily be extended to consider arbitrary planar arrangements of antennas [1-2] instead of the simple linear configuration employed here. The transmission radiation pattern of Eq. (1) can be tailored using common beamforming synthesis techniques [1-2] that rely on controlling the excitation amplitude $w_p$, the phases $\psi_p$, and, in this scheme, also the phases $\varphi_{mp}$. In particular, manipulating $\varphi_{mp}$ is advantageous because it requires phase-shifters operating at the low-frequency $f_m$ and avoids locating them in the path of the transmitted/received signals, which significantly reduces the impact of the phase shifters loss and other effects in the overall performance of the array. We note that this schematic resembles the reciprocal architecture of phased-arrays using the local-oscillator phase-shifting approach [77-79].

Consider now the phased-array antenna operating in reception, as illustrated in Fig. 1c. Using the array factor employed before, the radiation pattern of the antenna operated in reception, $\mathbf{E}_r$, can be computed as

$$\mathbf{E}_r(\theta,\phi) = \mathbf{E}_{ant}(\theta,\phi) \sum_{p=1}^{P} w_p e^{j(\psi_p - \varphi_{mp})}. \tag{2}$$

We stress that the array receives waves coming from free-space that oscillates at $f_0 + f_m$ and down-convert them to guided waves at $f_0$ ($n = -1$), which enforces a change of sign in the phases $\varphi_{mp}$ with respect to the transmission case. A simple analysis of Eqs. (1)-(2) reveals that appropriately controlling the phases $\psi_p$ and $\varphi_{mp}$ permits to shape drastically different radiation patterns in transmission and reception by taking advantage of available and very-well developed beamforming synthesis techniques [1-5]. For instance, if all antenna elements are fed with the same phase at $f_0$, i.e., constant $\psi_p \ \forall p$, the spatial angles of maximum transmission and reception of energy will always be opposite, i.e., $(\theta_t^{max}, \phi_t^{max}) = (-\theta_r^{max}, -\phi_r^{max})$ where the subscripts '$r$' and '$t$' denotes reception and transmission, respectively. Even greater flexibility and exciting functionalities can be obtained by also controlling the phases of the elements at $f_0$ ($\psi_p$), including tuning the spatial angle of maximum transmission (reception) in real-time while simultaneously preventing any reception (transmission) of energy from (to) that direction.

**III. Exploiting symmetries in nonlinear resonant antennas**
We introduce here an approach to achieve very efficient frequency conversion between spatial and guided waves in nonlinear resonant antennas based on exploiting even and odd symmetries in the structure through a feedback mechanism. The resulting antennas exhibit the desired nonreciprocity in phase, following the scheme shown in Fig. 1a.

Consider a resonant, linear, half-wavelength antenna, such as a dipole or a patch antenna [2], with a resonant frequency $f_r$ and a bandwidth $\Delta f$. This type of structures supports surface currents (electric fields) with an even (odd) symmetry with respect to the center of the antenna, as illustrated in Fig. 2a. Such symmetries can be further manipulated by simultaneously exciting the antenna from two symmetrical ports. The equivalent circuit of such device is composed of two identical resonators coupled through a resistor $R_{rad}$ that models the antenna radiation to free-space. When the exciting signals are in-phase, the symmetric (even) mode of the antenna is excited thus preventing any current flowing on $R_{rad}$ and, in turn, any radiation to free-space. The surface currents and electric field induced along the structure in this case exhibit odd and even symmetries, respectively. When the exciting signals are 180 degrees out-of-phase, the antisymmetric (odd) mode is excited. Then, currents can flow through $R_{rad}$ and the total radiation to free-space is maximized. For the sake of simplicity, we have neglected the presence of dissipation loss in this simple model, but it can easily be included by incorporating additional resistors in the circuit.

We propose to exploit these even/odd modes to implement electromagnetic resonances for spatial and guided waves that will enable very efficient frequency conversion between them. To do so, we first feed the two ports of the antenna from the same input line, creating a loop that serves as a feedback mechanism. And second, we will include a variable capacitor on each resonator as a tuning element. The equivalent circuit of the resulting antenna is shown in Fig. 2b. The varactors are time-modulated following

$$C_1(t) = C_0 \left[ 1 + \Delta_m \cos(2\pi f_m t + \varphi_m) \right], \tag{3}$$

$$C_2(t) = C_0 \left[ 1 + \Delta_m \cos(2\pi f_m t + \varphi_m + \pi) \right], \tag{4}$$

where $\Delta_m$ is the modulation index, $C_0$ denotes the average capacitance, and a phase difference of 180 degrees has been imposed between the signals that modulate each varactor. The time-modulated resonators create nonlinear harmonics on the circuit. For a given harmonic, the signals generated on both resonators have identical amplitude and a relative phase difference of $n\pi$, being $n \in \mathbb{Z}$ the harmonic order, that appears due to the different initial phases of the time-modulated capacitors. In general, the amplitude of each harmonic depends on a non-trivial manner with the antenna structure and the scheme applied to modulate the resonators, i.e., the modulation frequency and modulation index ($f_m, \Delta_m$), as described in Appendix B.

In order to investigate the linear response of the proposed antenna configuration, Fig. 3 explores the phase of the reflection coefficients in the absence of time-modulation for the specific case of a patch antenna at microwaves (see inset). As further discussed later in the text, the patch has been modified following the guidelines detailed above. In the figure, the red line denotes the phase of the reflection coefficient when the antenna is excited with a plane wave coming from free-space (spatial wave). The blue line shows the phase of the reflection coefficient when the structure is excited with a microstrip port (guided wave). Abrupt changes in the phases reveal the presence of strong resonances for guided and spatial waves at $f_0$ an $f_r$, respectively. Note that for the case of the guided wave the phase of the reflection coefficient undergoes several -180º to +180º transitions around the resonant frequency. For clarity, unwrapped phases are shown in Fig. 3. This double resonance behavior can be enforced in any resonant antenna that is adequately fed from its sides.

The response of the proposed nonlinear antenna operating in transmission, i.e., when the input port is excited with a signal oscillating at $f_0$, can be analyzed considering the even and odd symmetries that the excited harmonics enforce on the circuit (see Appendix B). In case of odd harmonics ($n = \pm 1, \pm 3$ ...), the signals generated on the resonators at $f_0 + nf_m$ are 180 degrees out-of-phase and excite the antisymmetric mode of the structure. The equivalent circuit of this scenario is shown in Fig. 2c (top), where the top and the bottom electric networks characterize the antenna seen from its left and right sides, respectively. For a given harmonic, the signals generated on each resonator have been modeled using a circuit-dependent current source with identical amplitude and opposite phase. Taking advantage of the circuit symmetry, it can be shown that the currents flowing on each network of Fig. 2c (top) are out-of-phase and that they destructively interfere when propagating back to the input port. Therefore, odd harmonics can only propagate towards the antenna. At this point, one can take advantage of the resonator filtering behavior to strongly favor the generation of only one desired harmonic. Specifically, if the frequency of such harmonic is equal to the resonant frequency of the antenna, i.e., $f_0 + nf_m = f_r$, the input impedance of both networks shown in Fig. 2c (top) will be purely real and equal to $R_{rad}/2$. As a result, the antenna will be efficiently excited, and the energy will be radiated to free-space at $f_0 + nf_m$ with a phase $n\varphi_m$. To ensure maximum conversion efficiency between these frequencies, it is important to simultaneously adjust the linear response of the device at $f_0$ and $f_0 + nf_m$. This condition is equivalent to the phase-matching requirement usually employed in nonlinear optics [55] and corresponds to a photonic transition between guided and spatial waves ($f_0 \rightarrow f_0 + f_m$) in the example of Fig. 3. We remark that modifying the values of the varactors simultaneously affects the two resonances supported by the structure. In case that the frequency of the odd harmonic is not at the antenna resonance, i.e., $f_0 + nf_m \neq f_r$, and assuming that the modulation frequency is larger than the bandwidth of the resonator ($f_m \gg \Delta f$), the input impedance of the networks shown in Fig. 2c (top) will be mostly reactive. Such input impedance tends to be a short circuit as the frequency of the harmonic is far away from the antenna resonance. As a result, this type of odd harmonics cannot deliver power to any resistive load, being mostly reactively and thus barely excited. Let us analyze now the response of the device for even harmonics ($n = \pm 2, \pm 3$ ...). In this case, the signals generated on the resonators at $f_0 + nf_m$ will be in phase and will excite the symmetric mode of the structure. The equivalent circuit of this scenario is shown in Fig. 2c (bottom), where it is evident that electrical currents cannot flow through the radiation resistance $R_{rad}$ and therefore the antenna cannot radiate. Following similar arguments as above, it can be shown that when the frequency of the generated harmonics is equal to the antenna resonant frequency, i.e., $f_0 + nf_m = f_r$, these signals will constructively interfere and propagate towards the input port. At other frequencies, when $f_0 + nf_m \neq f_r$, the generated nonlinear harmonics will be loaded by a mostly reactive impedance and thus they will not be strongly excited. We note that even harmonics might exist in the antenna even though they cannot efficiently outcouple to free-space or to the feeding network.

Consider now the antenna operated in reception. It receives energy at the resonant frequency $f_r$ that excites the antisymmetric (odd) mode of the structure. The energy cannot be guided to the output port at $f_r$ due to a destructive interference between the signals coming from both sides of the antenna. Instead, the received power is trapped in the structure thus favoring the generation of nonlinear harmonics in the time-modulated resonators. Again, one can analyze the antenna response taking advantage of the even/odd symmetries imposed on the structure by the generated harmonics. On one hand, even harmonics excite the antisymmetric mode that prevents them to be guided towards the output port. Besides, they cannot be re-radiated to free-space and thus are barely excited. On the other hand, odd harmonics will excite the symmetric (even) mode of the structure. The signals generated on the resonators will be in-phase and will constructively interfere. However, efficient frequency conversion will occur only between guided and spatial waves that fulfill the phase-matching conditions discussed above and illustrated in Fig. 3 for the first odd harmonic. Therefore, if a given antenna is designed to be feed at $f_0$ and radiate at $f_0 + nf_m = f_r$ (i.e., $f_0 \rightarrow f_0 + nf_m$) with phase $n\varphi_m$, then it will receive energy at $f_r = f_0 + nf_m$ and will guide it towards the output port at $f_0$ (i.e., $f_0 + nf_m \rightarrow f_0$) with phase $-n\varphi_m$. Both frequency conversion processes will provide similar efficiency when $f_m \ll f_0$ (see Appendix B).

We emphasize that the proposed approach is general and can be applied to realize nonlinear resonant antennas with nonreciprocal phase response across the electromagnetic spectrum. In all cases, the transmission and reception of energy involve opposite odd harmonics $(n, -n)$ that provide similar conversion efficiencies and opposite phase response.

### IV. Time-modulated patch antennas

To demonstrate the proposed antenna concept, we have chosen to modulate the response of a common patch antenna operating at 2.4 GHz. The resulting structure, described in Fig. 4, exhibits tunable nonreciprocity in phase versus the phase of the modulation frequency. In this specific implementation, the RF signal ($f_0 = 2.09$ GHz) flows along microstrip lines printed on the top of a board to feed a patch antenna from two sides whereas modulation signals ($f_m = 310$ MHz) flow through two coplanar waveguides (CPWs) located in the ground plane. Each CPW is loaded with a shunt varactor and is connected to the top microstrip line through a via-hole (see Appendix C). Fig. 4a shows a photograph of the manufactured prototype. The operation principle of the antenna follows the scheme detailed above. At $f_0$, the patch antenna is fed simultaneously from two lateral sides to enforce even symmetry at its center (see Fig. 4b top). As a result, the fundamental mode of the patch cannot be excited, and the energy is simply reflected back to the microstrip lines. The varactors are located at roughly $\lambda_g/4$ (where $\lambda_g$ is the guided wavelength) from the patch center to create a strong interaction at frequency $f_0$ for the guided waves and to enforce strong coupling between the even and odd modes. Upon time-modulation, the varactors generate nonlinear harmonics at frequencies $f_0 \pm nf_m$, with $n \in \mathbb{Z}$. Most energy is up-converted to the desired nonlinear harmonic, $f_0 + f_m = 2.4$ GHz, because at this frequency the patch antenna resonates, exhibits an input impedance that is real, and the required phase-matching conditions are fulfilled. Other nonlinear harmonics are barely excited. We emphasize that a phase difference of 180° has been imposed between the modulation signals that control the two varactors, enforcing that the first odd harmonic signals generated on the patch resonator are out of phase and excite the odd symmetric mode of the antenna at $f_0 + f_m$, as illustrated in Fig. 4b (bottom) using numerical simulations. When operated in reception, the time-modulated antenna receives power at $f_0 + f_m$, down-converts it to $f_0$, and delivers it to the input port at $f_0$.

The response of the proposed antenna concept has been investigated using a fabricated prototype. We have measured its linear and nonlinear behavior and then studied the transmission and reception response of the patch in an anechoic chamber. Fig. 4c shows the excellent matching of the fabricated prototype at the design frequency $f_0 = 2.09$ GHz. Fig. 4d depicts the normalized spectrum of the power reflected by the antenna when it is excited at 2.09 GHz, confirming that the generated harmonics that are reflected back to the feeding network carry negligible power (at least 43 dB lower than the excitation signal). Fig. 4e-f show the power spectrum of the antenna operating in transmission and reception measured at the broadside direction. Results are normalized with respect to the response of a similar non-modulated patch antenna employed for reference purposes (see Appendix C and D). In Fig. 4e the device is excited at $f_0 = 2.09$ GHz. Most energy is efficiently converted to $f_0 + f_m = 2.4$ GHz and radiated, a process that exhibit a power loss of just 2.7 dB. Note that 0 dB loss would correspond to a transmitted power identical to the one of the reference unmodulated antenna. We attribute the power loss to the following damping mechanisms (i) excitation of unwanted nonlinear harmonics within the antenna structure; and (ii) dissipation in the equivalent resistors of the varactors. In Fig. 4f the antenna receives waves oscillating at $f_0 + f_m = 2.4$ GHz and efficiently down converts them to $f_0 = 2.09$ GHz. In all cases, our measured data show that the power transferred to undesired harmonics is at least 30 dB lower than those carried by desired signals. The nonreciprocal response of the antenna appears due to the different phases imparted to transmitted $(f_0 \rightarrow f_0 + f_m)$ and received $(f_0 + f_m \rightarrow f_0)$ signals, as demonstrated in Fig. 4g. Controllable nonreciprocity arises because the phase of transmitted signals follows the phase of the modulating signal $\varphi_m$, whereas the phase of the received signal follows the opposite phase, i.e., $-\varphi_m$. As expected, the radiation pattern of the antenna is identical in transmission and reception (see Appendix C).

## V. Nonreciprocal phased-array antennas

To demonstrate phased-array antennas exhibiting drastically different transmission and reception radiation patterns, one can exploit the multiple degrees of freedom provided by Eqs. (1)-(2). We have designed a two-element antenna array that maximize the isolation between transmission and reception at the broadside direction. The structure is composed of two patch antennas as the one described in Fig. 4 and a feeding network that excites them with 90° phase difference at $f_0 = 2.09$ GHz by simply using a slightly longer microstrip line to feed one of the antenna elements (see Fig. 5a). In addition, the modulation signals that control one patch have a phase difference of $\varphi_m$ with respect to those that control the second one, which is achieved in practice using a phase-shifter at $f_m = 310$ MHz. Therefore, one antenna element (left) acts as a reference and imparts a phase of $\psi_1 = 0°$ to the transmitted/received waves, whereas the other element (right) imparts a phase composed of $\psi_2 = -90°$ plus the nonreciprocal contribution from the modulation frequency phase $\varphi_m$. During the transmission of energy (Fig. 5a, top), the phases of the left and right antenna elements are given by $[0°, -90° + \varphi_m]$. Maximum radiation can be obtained by setting $\varphi_m = 90°$ because it enforces that the two antennas are in phase and provide a constructive interference at broadside [1-2]. During the reception of energy (Fig. 5a, bottom), the phases of the left and right antenna elements are $[0°, -90° - \varphi_m]$. Setting $\varphi_m = 90°$ enforces that the two antennas are out of phase and conform a destructive interference pattern at broadside. Table 1 presents an overview of the phases exhibited by the left and right elements of the antenna array when it operates in transmission or reception for several values of the phase $\varphi_m$. By appropriately controlling $\varphi_m$ it is possible to tailor the phase profiles exhibited by the array in reception and transmission, achieving strong nonreciprocal responses and enabling reconfigurable capabilities. To verify that this is indeed the case, a prototype has been fabricated and tested (see Fig. 5b and Appendix C). Fig. 5c shows the power transmitted by the two-element array at 2.4 GHz $(f_0 \rightarrow f_0 + f_m)$ normalized with respect to the power transmitted by an unmodulated array employed for reference (see Appendix C). Results, measured in an anechoic chamber at the broadside direction, are plotted versus the phase $\varphi_m$. When $\varphi_m = 90°$, both patches are in phase and the power radiated by the antenna array is maximum, exhibiting a power loss of just 3.5 dB. The inset shows the power spectrum of the transmitted waves, confirming that very little power (<-30 dB) has been transferred to undesired harmonics. When $\varphi_m = 270°$, the antenna elements are out of phase and the power radiated is minimum and 44 dB lower than the one from the reference antenna. Fig. 5d shows the antenna response when receiving electromagnetic waves oscillating at 2.4 GHz. When $\varphi_m = 90°$, the radiating elements are out of phase and the power received by the antenna is minimum and 44.3 dB lower than the one from the reference non-modulated array. The inset shows the power spectrum of the received energy. Our measured data confirms strong nonreciprocal transmission of energy, with an isolation level over 40 dB. When $\varphi_m = 270°$, the antenna elements are in phase and the power is adequately received and down-converted to 2.09 GHz. Compared to the non-modulated reference antenna array, a power loss of 3.9 dB has been measured. Note that the transmitted power is minimum in this case and a strong nonreciprocity over 40 dB is also obtained, favoring in this case the reception of energy. Fig. 5e shows the measured antenna radiation diagrams (E-plane) when $\varphi_m = 90°$ and $\varphi_m = 270°$. These results confirm that a drastically different radiation pattern in transmission and reception can be obtained. In addition, it shows that it is easy to favor the reception or transmission of energy, therefore, controlling the strength of the nonreciprocal response by simply manipulating the phase of a low-frequency signal. We note that reciprocal responses in amplitude for transmission and reception can be found imposing a phase of $\varphi_m = 0°$ or $\varphi_m = 180°$, as indicated by Fig. 5c-d and illustrated in Fig. 6. The response of the antenna array versus the modulation frequency, modulation index, and the DC biasing voltage applied to the varactor is shown in Fig. 7. Finally, Fig. 8 plots measured radiation patterns of the proposed array operating in transmission at 2.4 GHz for various values of the phase $\varphi_m$. As expected, in addition to nonreciprocal responses, time-modulated phased-array antennas exhibit beam scanning capabilities as those provided by common phased-array antennas [1-2].

It should be emphasized that the proposed nonreciprocal phased-array antenna paradigm can be further applied to independently control same-frequency transmission and reception patterns exploiting the degrees of freedom provided by Eqs. (1)-(2). The resulting arrays will be able to dynamically manipulate the level of nonreciprocity, favor transmission/reception at will, and incorporate advanced algorithms to synthesize complex radiation patterns and implement many functionalities. For instance, let us consider a linear array composed of 1x8 antenna elements (similar to the one described in Fig. 4) that are simultaneously fed by RF signals oscillating at $f_0 = 2.09$ GHz with identical phase (i.e., equal $\psi_p \, \forall p$). Numerical simulations shown in Fig. 9 confirm that radiation and transmission patterns follow an opposite behavior, i.e., maximum transmission and reception appear at opposite elevation angles ($\theta_t = -\theta_r$). Even more sophisticated radiation patterns and tailored nonreciprocal responses can be obtained by simultaneously controlling $\psi_p$ and $\varphi_m$.

## VI. Conclusions

We have proposed and demonstrated the concept of nonreciprocal phased-array antennas able to exhibit drastically different radiation patterns in transmission and reception. The underlying mechanism is based on relating states associated to spatial and guided waves in time-modulated antennas using photonic transitions and on taking advantage of the photonic Aharonov-Bohm effect to control in a nonreciprocal manner the phases of the waves that are transmitted and received by each antenna element. This platform adds an extra degree of flexibility to smart antennas: it enables exciting possibilities to boost the performance of communication and sensing systems, and open new opportunities to deal with jamming signals and strong interferences in electromagnetically crowded scenarios as well as in thermodynamics and energy harvesting. The fundamental building block of the array is a time-modulated resonant antenna that provides tunable nonreciprocity in phase when operated in transmission or in reception. We have introduced a general approach to efficiently design such nonlinear structures based on exploiting even and odd symmetries imposed on nonlinear harmonic frequencies. Measured results from a two-element array prototype based on time-modulated patch antennas confirm that (i) different patterns in transmission and reception can be achieved at the same frequency, showing isolation levels over 40 dB at desired directions; (ii) such patterns can be tailored with the phase of low-frequency modulation signals; and (iii) the process is very efficient, with reduced losses between 3 and 4 dB. Note that this is a first prototype of its kind that has not been fully-optimized. Therefore, we expect that efficiency can be increased even further in the near future.

We would like to emphasize that our approach to design nonreciprocal resonant antennas is relatively simple, very efficient, and can be applied to realize different resonant devices across the electromagnetic spectrum. For instance, they can be constructed from RF to micro and millimeter waves using tuning elements such as varactors or micro-electro-mechanical systems (MEMS) [81], from terahertz to infrared frequencies using graphene exploiting its ultra-fast field effect [82-84], and from infrared to optical frequencies using doped semiconductors [85]. Alternative antenna designs might allow it to operate with higher odd nonlinear harmonics, thus enabling additional flexibility to control the frequencies involved in the nonlinear process; use other resonant antennas such as dipoles, slots, or loops; simplify the antenna excitation using various coupling mechanisms; and even enable nonreciprocal dual-polarized responses.

Finally, nonreciprocal phased-array antennas can be extended from the linear configuration explored here to large planar arrangements with very high directivity and multi-beam responses. Such arrays can readily take advantage of smart antenna algorithms and beamforming synthesis techniques to implement well-known real-time applications, like simultaneously tracking several targets, but also novel functionalities that might require an independent control of the transmission and reception radiation patterns, including enhancing the capacity of a MIMO channel and avoiding strong jamming signals while maintaining the communication link. To this purpose, the resulting phased-arrays need a tunable feeding network to control the phases of the modulation signals (as schematically shown in Fig. 1b-c) and another one to manipulate the phases of the signals that feed the radiation elements, as usually done in common phased-array antennas. We envision that this technology will lead to a new generation of nonreciprocal smart antennas with wide applications in radar, sensing, and communication systems.

**Acknowledgement**
This work was supported by the National Science Foundation with CAREER Grant No. ECCS-1749177. A. Alvarez-Melcon acknowledges support from grants PRX18/00092, TEC2016-75934-C4-4-R, and 19494/PI/14 of MECD, Spain. Authors wish to thank Roger Rogers Corporation for the generous donation of the dielectrics employed in this work. In addition, authors wish to thank Mr. James Do (University of California, Davis) for help with the measurements and fruitful discussions and to Prof. X. Liu, Prof. J. Gu, Prof. Branner and Prof. N. C. Luhmann (University of California, Davis) for providing access to the equipment required to carry out this work.

**Appendix A - Numerical simulations**
Linear numerical simulations were carried out using the commercial software Ansys High Frequency Structure Simulator (HFSS). Nonlinear numerical simulations were carried out using the commercial software Keysight Advanced Design System (ADS). In all cases, results from numerical simulations agree very well with experimental data and are not shown for the sake of clarity.

**Appendix B - Theoretical analysis**
Let us consider the linear circuit of Fig. 2a. We model the coupling between the port and the resonators using common admittance inverters (see Fig. B1). Applying Kirchhoff's current law to the nodes of the resulting network, the following system of linear equations can be derived

$$\boldsymbol{I}_S = \left[\overline{\boldsymbol{G}}_P + \overline{\boldsymbol{Y}}_{inv} + \overline{\boldsymbol{Y}}_p + \overline{\boldsymbol{G}}_{rad}\right] \cdot \boldsymbol{V}, \tag{A1}$$

where

$$\boldsymbol{I}_S = \begin{pmatrix} I_{p1} \\ 0 \\ 0 \\ I_{p2} \end{pmatrix}, \overline{\boldsymbol{G}}_P = \begin{pmatrix} G_p & 0 & 0 & 0 \\ 0 & 0 & 0 & 0 \\ 0 & 0 & 0 & 0 \\ 0 & 0 & 0 & G_p \end{pmatrix}, \overline{\boldsymbol{Y}}_{inv} = j\begin{pmatrix} 0 & J_p & 0 & 0 \\ J_p & 0 & 0 & 0 \\ 0 & 0 & 0 & J_p \\ 0 & 0 & J_p & 0 \end{pmatrix}, \overline{\boldsymbol{Y}}_p = \begin{pmatrix} 0 & 0 & 0 & 0 \\ 0 & Y_{p1} & 0 & 0 \\ 0 & 0 & Y_{p2} & 0 \\ 0 & 0 & 0 & 0 \end{pmatrix}, \tag{A2}$$

$$\overline{\boldsymbol{G}}_{rad} = \begin{pmatrix} 0 & 0 & 0 & 0 \\ 0 & G_{rad} & -G_{rad} & 0 \\ 0 & -G_{rad} & G_{rad} & 0 \\ 0 & 0 & 0 & 0 \end{pmatrix}, \boldsymbol{V} = \begin{pmatrix} V_{p1} \\ V_1 \\ V_2 \\ V_{p2} \end{pmatrix}$$

are matrixes that denote excitation coming from ports 1 and 2 ($I_S$); the port admittances ($\overline{G}_P$), with $G_p = 1/R_p$; the coupling between the ports and the resonators ($\overline{Y}_{inv}$), modelled through admittance inverters; the admittance of the resonators ($\overline{Y}_p$), with $Y_{p1} = Y_{p2} = j\omega C + \frac{1}{j\omega L}$; the radiation conductance ($\overline{G}_{rad}$), with $G_{rad} = 1/R_{rad}$; and the voltages on the nodes of the network ($V$), following the scheme of Fig. B1. In case that ports 1 and 2 are identically excited (even mode), it is easy to show that $I_{p1} = I_{p2}$ and $V_{p1} = V_{p2}$, which leads to $V_1 = V_2$. As a consequence, the current flowing through the radiation resistor $I_{rad} = (V_1 - V_2)/R_{rad}$ is strictly zero and there is no radiation towards free-space. If the excitation from ports 1 and 2 has the same amplitude but opposite phase (odd mode), then $I_{p1} = -I_{p2}$ and $V_{p1} = -V_{p2}$ which leads to $V_2 = -V_1$. Then, the current flowing through the resistance is maximum at the resonant frequency and the energy is radiated to free space. We remark that this formulation is rigorous, and that no approximation has been introduced so far. Let us now time-modulate the varactors following the scheme of Eqs. (3)-(4). Nonlinear harmonics are generated in each resonator, leading to the equivalent network shown in Fig. B2. Applying Kirchhoff's current law to this circuit and considering that a finite number $N_h$ of nonlinear harmonics have been generated lead to a system of linear equations very similar to the one of Eq. (A1). The key difference is that each element of the matrixes now becomes a submatrix with size $N_h \times N_h$ that takes the presence of the different harmonics and their interaction into account. Specifically,

$$\begin{pmatrix} I_{p1} \\ 0 \\ 0 \\ I_{p2} \end{pmatrix} = \left[ \begin{pmatrix} \overline{G}_p & j\overline{J}_p & \overline{0} & \overline{0} \\ j\overline{J}_p & \overline{0} & \overline{0} & \overline{0} \\ \overline{0} & \overline{0} & \overline{0} & j\overline{J}_p \\ \overline{0} & \overline{0} & j\overline{J}_p & \overline{G}_p \end{pmatrix} + \begin{pmatrix} \overline{0} & \overline{0} & \overline{0} & \overline{0} \\ \overline{0} & \overline{Y}_{p1} + \overline{G}_{rad} & -\overline{G}_{rad} & \overline{0} \\ \overline{0} & -\overline{G}_{rad} & \overline{Y}_{p2} + \overline{G}_{rad} & \overline{0} \\ \overline{0} & \overline{0} & \overline{0} & \overline{0} \end{pmatrix} \right] \begin{pmatrix} V_{p1} \\ V_1 \\ V_2 \\ V_{p2} \end{pmatrix}, \quad (A3)$$

where $\overline{0}$ and $0$ are the zero matrix and vector, respectively; $\overline{G}_p = G_p \overline{T}$, $\overline{J}_p = J_p \overline{T}$, and $\overline{G}_{rad} = G_{rad} \overline{T}$ being $\overline{T}$ the identity matrix; $I_{p1} = (\cdots \quad 0 \quad I_{p1} \quad 0 \quad \cdots)'$ and $I_{p2} = (\cdots \quad 0 \quad I_{p2} \quad 0 \quad \cdots)'$ represents the ports excitation at the fundamental frequency, being ' the matrix transpose; and $V_{p1}$, $V_{p2}$, $V_1$, and $V_2$ are vectors that model the harmonic voltages on ports 1 and 2 and on the nodes 1 and 2 of the network, respectively (see Fig. B2). The admittance submatrix of the time-modulated resonators including the coupling among all excited harmonics can be derived analytically using the theory developed in Refs. [86-87]. Then, the matrixes that characterize the time-modulated resonators 1 and 2 can be expressed as

$$\overline{Y}_{p1} = \begin{pmatrix} \ddots & \vdots & \vdots & \vdots & \vdots & \vdots & \reflectbox{$\ddots$} \\ \cdots & Y_r^{(-2)} & j\frac{\Delta_m C}{2}(\omega - \omega_m) & 0 & 0 & 0 & \cdots \\ \cdots & j\frac{\Delta_m C}{2}(\omega - 2\omega_m) & Y_r^{(-1)} & j\frac{\Delta_m C}{2}(\omega) & 0 & 0 & \cdots \\ \cdots & 0 & j\frac{\Delta_m C}{2}(\omega - \omega_m) & Y_r^{(0)} & j\frac{\Delta_m C}{2}(\omega + \omega_m) & 0 & \cdots \\ \cdots & 0 & 0 & j\frac{\Delta_m C}{2}(\omega) & Y_r^{(1)} & j\frac{\Delta_m C}{2}(\omega + 2\omega_m) & \cdots \\ \cdots & 0 & 0 & 0 & j\frac{\Delta_m C}{2}(\omega + \omega_m) & Y_r^{(2)} & \cdots \\ \reflectbox{$\ddots$} & \vdots & \vdots & \vdots & \vdots & \vdots & \ddots \end{pmatrix},$$

$$\overline{Y}_{p2} = \begin{pmatrix} \ddots & \vdots & \vdots & \vdots & \vdots & \vdots & \reflectbox{$\ddots$} \\ \cdots & Y_r^{(-2)} & -j\frac{\Delta_m C}{2}(\omega - \omega_m) & 0 & 0 & 0 & \cdots \\ \cdots & -j\frac{\Delta_m C}{2}(\omega - 2\omega_m) & Y_r^{(-1)} & -j\frac{\Delta_m C}{2}(\omega) & 0 & 0 & \cdots \\ \cdots & 0 & -j\frac{\Delta_m C}{2}(\omega - \omega_m) & Y_r^{(0)} & -j\frac{\Delta_m C}{2}(\omega + \omega_m) & 0 & \cdots \\ \cdots & 0 & 0 & -j\frac{\Delta_m C}{2}(\omega) & Y_r^{(1)} & -j\frac{\Delta_m C}{2}(\omega + 2\omega_m) & \cdots \\ \cdots & 0 & 0 & 0 & -j\frac{\Delta_m C}{2}(\omega + \omega_m) & Y_r^{(2)} & \cdots \\ \reflectbox{$\ddots$} & \vdots & \vdots & \vdots & \vdots & \vdots & \ddots \end{pmatrix},$$

(A4)

where only 4 nonlinear harmonics have been shown for the sake of brevity and $Y_r^{(k)} = jC(\omega + k\omega_m) + \frac{1}{jL(\omega+k\omega_m)}$ denotes the response of the resonator at the frequency of the $k^{\text{th}}$ harmonic. It is important to note that the form of the matrixes in Eq. (A4) permits us to understand the time-modulated circuit as a network of coupled nonlinear resonators [86-87]. Specifically, each diagonal element can be interpreted as a new resonator that appears at the nonlinear harmonic frequency $\omega + k\omega_m$. Besides, the off-diagonal elements in the matrixes show that the new resonators are coupled following an *in-line topology*, which is very common in the field of microwave filters [88]. That means that, within a physical resonator, a given nonlinear harmonic $k$ can only couple to the adjacent nonlinear harmonics, i.e., to $k + 1$ and to $k - 1$. Then the coupling between two arbitrary nonlinear resonators ($n,k$, with $k > n$), can be modeled in each physical resonator using impedance inverters, yielding

$$\text{Physical resonator 1} \begin{cases} J_{R1}^{(k,n)} = \frac{\Delta_m C_0}{2}(\omega_0 + k\omega_m) \rightarrow \text{up conversion} \\ J_{R1}^{(n,k)} = \frac{\Delta_m C_0}{2}(\omega_0 + n\omega_m) \rightarrow \text{down conversion} \end{cases}$$

$$\text{Physical resonator 2} \begin{cases} J_{R1}^{(k,n)} = (-1)^{(k-n)}\frac{\Delta_m C_0}{2}(\omega_0 + k\omega_m) \rightarrow \text{up conversion} \\ J_{R1}^{(n,k)} = (-1)^{(n-k)}\frac{\Delta_m C_0}{2}(\omega_0 + n\omega_m) \rightarrow \text{down conversion} \end{cases} \quad \textbf{(A5)}$$

where we highlight the sign difference that appears in the coupling between adjacent nonlinear harmonics in the first and second physical resonator. It arises due to the 180° phase difference between the signals that modulate the varactors (see Eqs. (3)-(4)). At this point, Eq. (A3) can be numerically solved. It will give very accurate results provided that an adequate number of nonlinear harmonics are included in the computation. In our numerical study, we achieve convergence using between 5 and 7 nonlinear harmonics. We have employed the commercial software Keysight Advanced Design System (ADS) to validate the accuracy of our results.

Our analysis above has demonstrated that nonlinear harmonics within a given physical resonator couple among themselves following an in-line topology. In addition, the symmetry of the circuit imposes that the amplitude of a given nonlinear harmonic of order $k$ generated in both physical resonators must be identical. However, their phase can be different. In case of even harmonics, the in-line topology enforces that the coupling to these harmonics will always be positive for both physical resonators. Therefore, even harmonics will excite the symmetric (even) mode of the structure, which combined with the admittance of the resonator $Y_r^{(k)}$ leads to the equivalent circuit shown in Fig. 2c (top). In case of odd harmonics, the in-line topology imposes that the excitation of the second physical resonator is 180 degrees out-of-phase with respect to the one generated on the first physical resonator. As a result, odd harmonics will excite the antisymmetric (odd) mode of the structure, which combined with the admittance of the resonator $Y_r^{(k)}$ leads to the equivalent circuit shown in Fig. 2c (bottom). Finally, it should be emphasized that the amplitude of the coupling mechanism is very similar for up/down conversion processes provided that $\omega_m \ll \omega_0$ (see Eq. (A5)).

**Appendix C - Description of the fabricated antennas**
Details and dimensions of the fabricated nonreciprocal antennas can be found in Figs. C1 and C2. In all cases, a substrate Roger Corporation laminate RT/duroid 5880 with a thickness, permittivity, and tangent loss of $h = 1.575$ mm, $\varepsilon_r = 2.2$ and $\tan\delta = 0.0009$, respectively, has been used. The plated via-holes employed to connect the microstrip lines and patch antenna with the CPWs located in the ground plane have a diameter of 0.4mm. Each CPW is linked to a via-hole through a varactor Skyworks SMV1233 and a lumped inductor TDK SIMID with 33nH (choke) connected in parallel (see Fig. C1). In addition, Figs. C3-C5 show the dimensions and details of the reference unmodulated patch antennas that serve as a reference.

**Appendix D - Experimental characterization**

Modulation signals oscillating at 310 MHz were generated with a signal generator Hewlett Packard E4433B and used to feed the CPWs of the nonreciprocal antennas. Each individual time-modulated patch antenna is controlled with two modulation signals that exhibit a phase difference of 180 degrees between them (see Fig. 4a), which is obtained in practice using phase shifters Mini-Circuits ZXPHS-431. In addition, the phase difference between the signals that control each patch antenna in the 2-element array ($\varphi_m$ in Fig.5a-b) was precisely determined with a phase shifter RF-Lambda RVPT0205MBC controlled with a DC source. The linear scattering parameters of the antennas were measured using a N5247A PNA-X Keysight vector network analyzer. To measure the nonlinear response of the antenna in reflection (Fig. 4d), the device was excited at 2.4 GHz through a directional coupler Krytar 1815 using an additional signal generator Hewlett Packard E4433B and the reflected waves were analyzed with a N9030A PXA Keysight signal analyzer. The transmission and reception properties of the fabricated antennas were tested in an anechoic chamber using a standard S-band horn antenna. To compute the transmission (reception) radiation diagram, the antenna under test (horn antenna) is fed at 2.09 GHz (2.4 GHz) using the signal generator and the propagating waves are received at 2.4 (2.09 GHz) by a vector network analyzer. To compute the nonlinear transmission/reception properties of the antenna, the same procedure as described before is followed but the propagating waves are received with the signal analyzer.

The power transmitted/received by the proposed time-modulated patch antennas when operated in the anechoic chamber has been normalized with respect to the one measured from standard patch antennas (see Appendix C and Figs. C3-C5). All antennas have been designed to provide identical gain at 2.4 GHz.

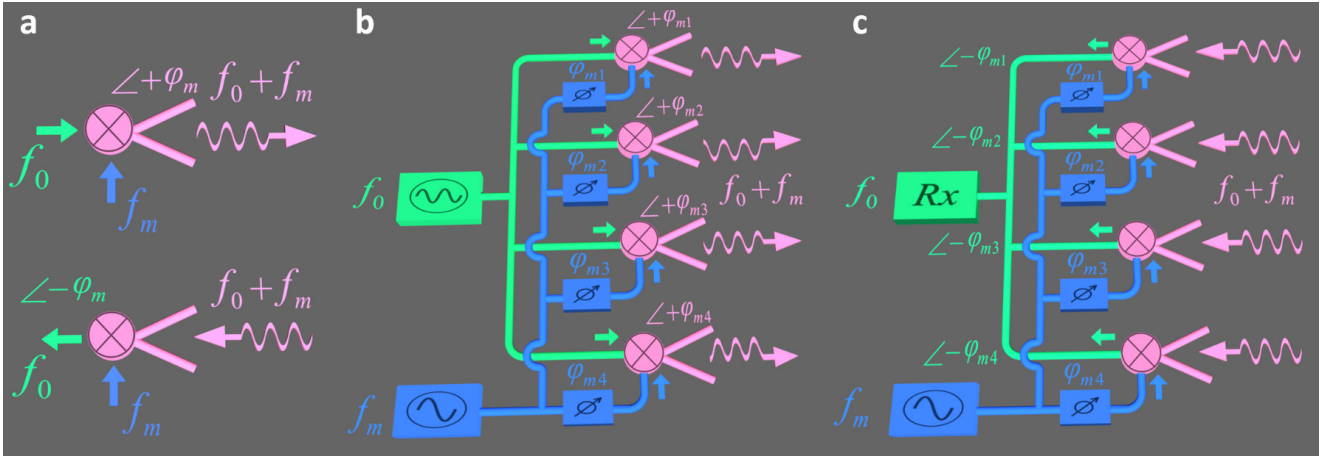

**Fig. 1. Architecture of nonreciprocal phased-array antennas**. **a**, Schematic of the proposed nonlinear resonant antenna (magenta) when its electromagnetic response is modulated with a signal with frequency $f_m$ and phase $\varphi_m$. In transmission (top) the antenna up-converts the exciting signal $f_0$ to $f_0 + f_m$ and radiates it towards free-space with a phase proportional to $+\varphi_m$. In reception (bottom) the antenna down-converts the incoming waves oscillating at $f_0 + f_m$ to $f_0$ with a phase proportional to $-\varphi_m$. **b**, **c**, Diagram of a nonreciprocal phased-array antenna operating in transmission **(b)** and in reception **(c)**. The device consists of feeding networks for the signal $f_0$ (green) and for the low-frequency modulation signal $f_m$ (blue), phase shifters operating at $f_m$ (blue) that are controlled by a computer, and time-modulated radiating elements (magenta) that transmit and receive electromagnetic waves oscillating at $f_0 + f_m$. The phase profile exhibited by the time-modulated array when operated in transmission and in reception is opposite. This schematic can be extended to consider arrays of any number of radiating elements arbitrarily arranged in space.

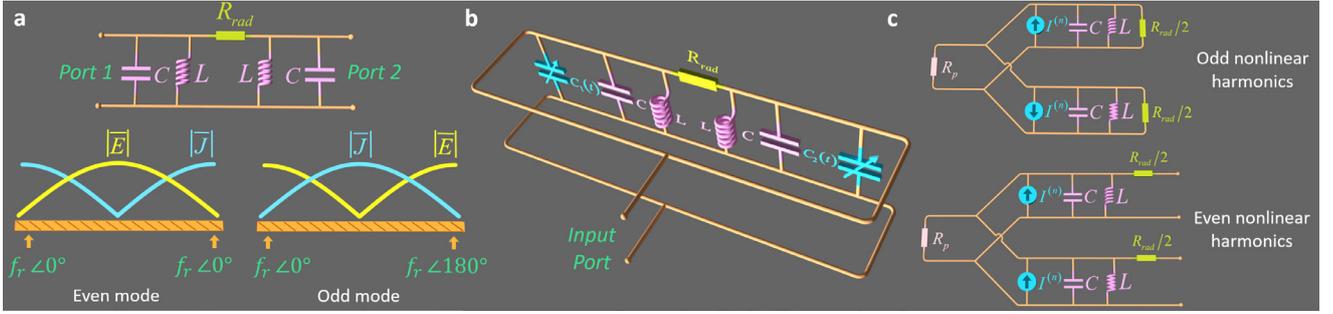

**Fig. 2. Even and odd symmetries in time-modulated resonant antennas**. **a**, Schematic representation of the surface current and electric field (magnitude) induced in a one-dimensional, half-wavelength, and linear antenna simultaneously fed from two ports at its resonant frequency $f_r$. The antenna supports a symmetric (even) mode when the two ports are excited in-phase and an anti-symmetric (odd) mode when the signals coming from the ports are 180 degrees out-of-phase. The structure can be characterized with an equivalent circuit composed of two resonators (LC tanks) coupled through a radiation resistance that models the radiation to free-space. **b**, Equivalent circuit of the proposed time-modulated resonant antenna. The structure is excited from a single input port that is connected in parallel with both sides of the antenna. The variable capacitors $C_1$ and $C_2$ are time-modulated with 180° phase difference following Eq. (3) and Eq. (4), respectively. **c**, Response of the proposed antenna at the nonlinear harmonic frequency $f_0 + nf_m$, where $n$ is the harmonic order and $f_0$ and $f_m$ are the excitation and modulation frequencies, respectively. Odd (even) harmonics impose an odd (even) symmetry in the structure, leading to the equivalent circuit shown in top (bottom) panel. $R_p$ denotes the port impedance and the amplitude of each harmonic is modeled with a current source $I^{(n)}$, as detailed in Appendix B.

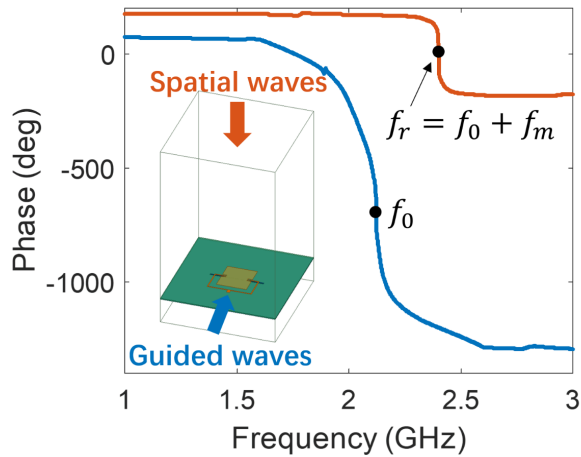

**Fig. 3. Phase of the reflection coefficient of an unmodulated resonant antenna for spatial (red line) and guided (blue line) waves.** The device is designed to exhibit two resonances: one at $f_r = f_0 + f_m$ provided by the even mode of the patch when it is excited by a plane wave coming from free-space (normal direction with respect to the structure); and another one at $f_0$ that appears for guided signals coming from the feeding network between the virtual open at the center of the patch center and the varactors located at each side of the structure. Further details about the antenna configuration are provided in Fig. 4.

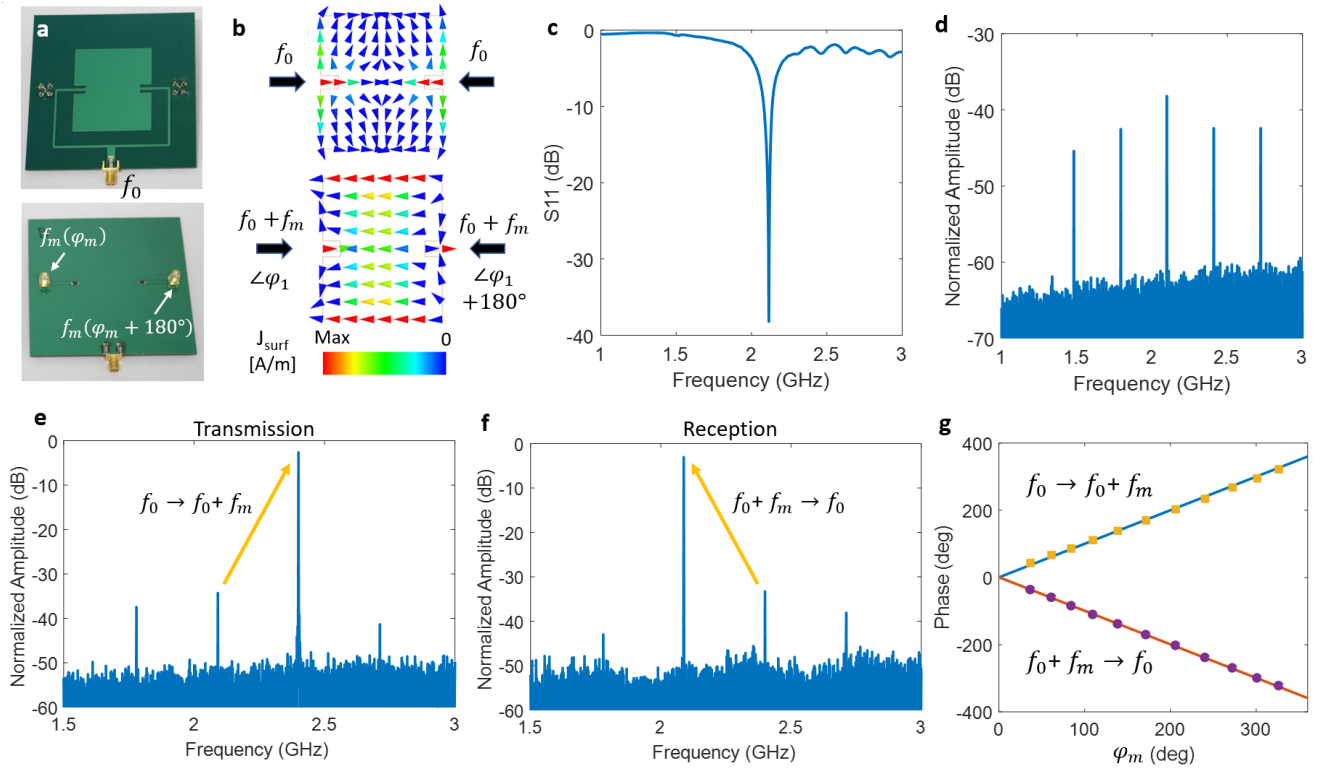

**Fig. 4: Time-modulated patch antenna with nonreciprocal phase response**. **a**, Photograph of a fabricated prototype detailing the RF signal $f_0$ on the top panel. Modulation signals oscillating at $f_m = 310$ MHz flow through coplanar waveguides located in the ground plane and have a phase difference of 180 degrees. The coplanar waveguides are loaded with a shunt varactor and a via-hole that connects with the feeding network of the patch. Modulation index is set to $\Delta_m = 0.29$. Further details are provided in Appendix C. **b**, Surface current induced in the patch antenna obtained through numerical simulations. At $f_0$, the antenna is simultaneously excited from two sides to enforce an even symmetry that prevents the excitation of the fundamental mode. The time-modulated varactors convert most energy to the nonlinear harmonic $f_0 + f_m$ that excites the patch antenna from both sides with 180 degrees phase difference between them. **c**, Measured reflection coefficient of the time-modulated antenna. **d**, Spectrum of the power reflected back to the input port when the antenna is excited at $f_0 = 2.09$ GHz. **e, f**, Spectrum of the power transmitted and received by the time-modulated antenna measured at the broadside direction in an anechoic chamber. The power is normalized with respect to the one of an unmodulated patch antenna employed as reference, as detailed in Appendix D. In **e** the antenna is excited at $f_0 = 2.09$ GHz and in **f** the antenna receives power oscillating at $f_0 + f_m = 2.4$ GHz. The yellow arrow indicates the direction of the frequency conversion. **g**, Measured (markers) and simulated (solid lines) phase of the signals transmitted ($f_0 \to f_0 + f_m$) and received ($f_0 + f_m \to f_0$) by the antenna versus the phase $\varphi_m$ of the modulating signals.

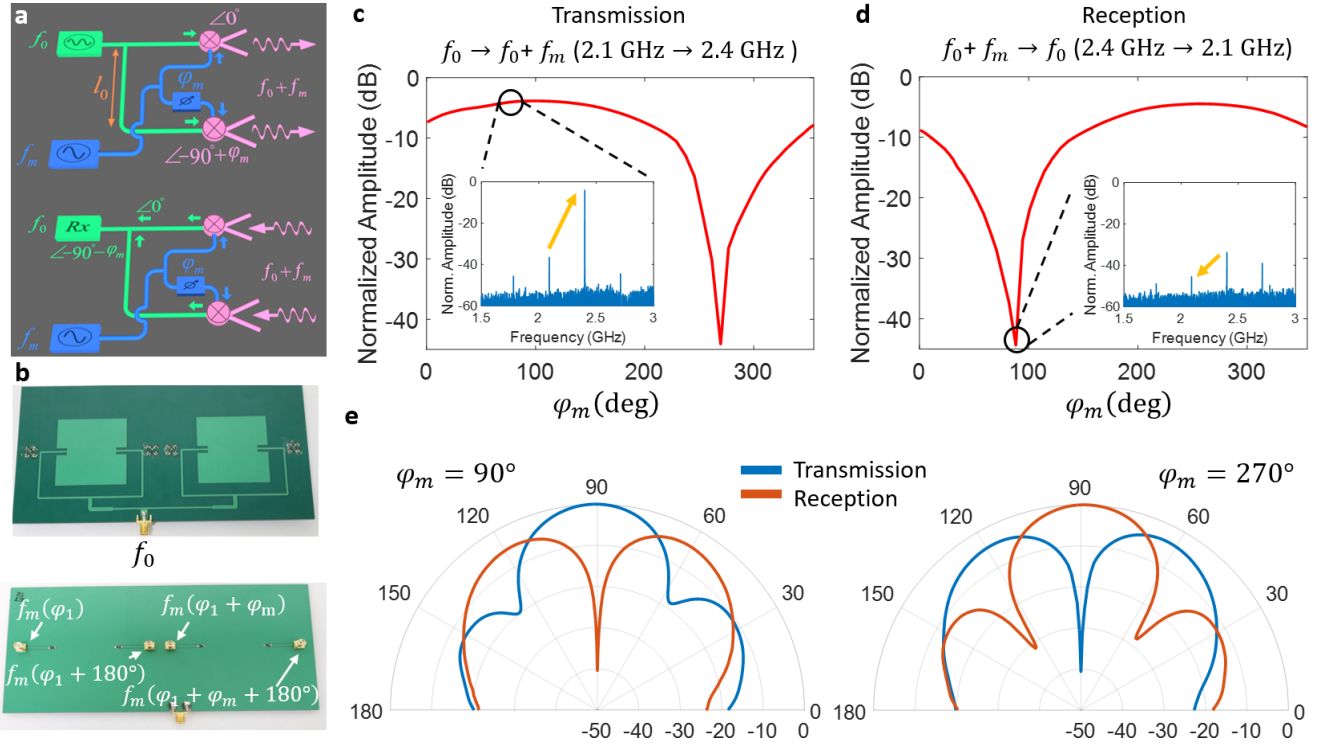

**Fig 5. Nonreciprocal phased array antenna composed of two time-modulated patch antennas**. **a**, Schematic of the array detailing its operation in transmission (top) and reception (bottom). **b**, Photograph of a fabricated prototype. The device is composed of two patches as the one described in Fig. 4 and a feeding network at $f_0$ that excites them with a phase difference of 90 degrees. The modulation signals that control each antenna have a phase difference of $\varphi_m$. **c**, Power transmitted at 2.4 GHz ($f_0 \rightarrow f_0 + f_m$) normalized with respect to the power transmitted by a reference unmodulated antenna, as detailed in Appendix D. Results, measured in an anechoic chamber at the broadside direction, are plotted versus the phase $\varphi_m$. The inset shows the transmitted power spectrum when $\varphi_m = 90°$. **d**, Power received by the antenna at 2.4 GHz that is subsequently converted to 2.09 GHz ($f_0 + f_m \rightarrow f_0$). Results, measured in an anechoic chamber at the broadside direction, are normalized with respect to the power received by a reference unmodulated antenna and are plotted versus the phase $\varphi_m$. The inset shows the received power spectrum when $\varphi_m = 90°$. **e**, Measured radiation diagram (E-plane) in dB of the antenna array operating in transmission (blue solid line) and reception (red solid line) when $\varphi_m = 90°$ (left panel) and $\varphi_m = 270°$ (right panel).

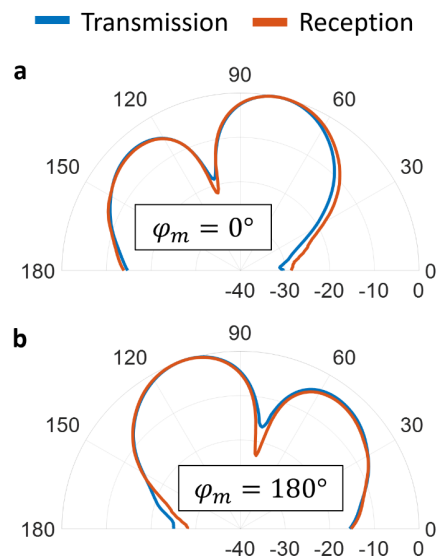

**Fig. 6. Measured transmission and reception radiation diagrams (E-plane) in dB of the nonreciprocal phased-array described in Fig. 5.** Results are plotted for the two values of the phase difference between the signals that modulate each patch antenna ($\varphi_m$) that yield reciprocal responses. (a) $\varphi_m = 0°$. (b). $\varphi_m = 180°$.

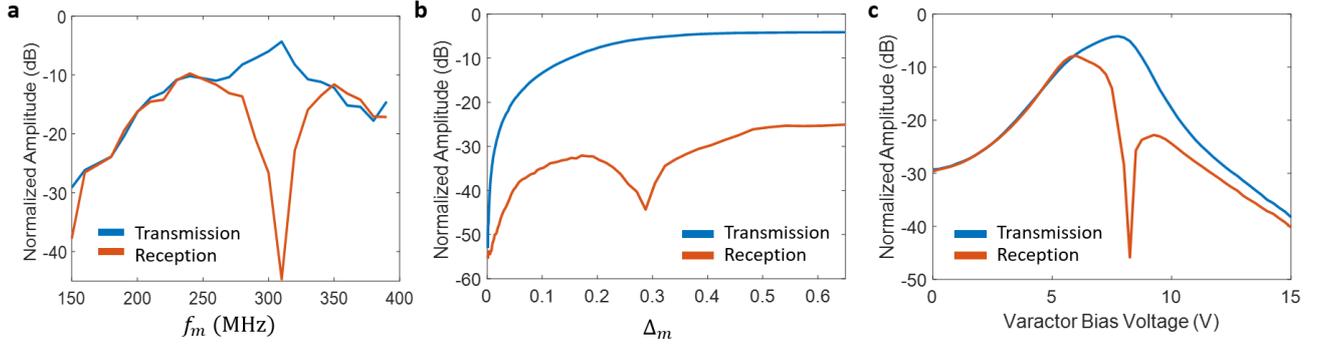

**Fig. 7. Measured transmission and reception properties of the nonreciprocal phased array described in Fig. 5 versus the features of the time-modulation scheme.** Results show the power transmitted and received by the antenna at the broadside direction measured in an anechoic chamber and normalized with respect to power of the standard 2-element antenna array described in Appendix C. Measured data is plotted versus (a) modulation frequency $f_m$, keeping the modulation index and DC varactor bias voltage to $\Delta_m = 0.29$ and $V_{DC} = 8.25\ V$, respectively; (b) modulation index $\Delta_m$, keeping the modulation frequency $f_m$ and DC varactor bias voltage to $f_m = 310$ MHz and $V_{DC} = 8.25\ V$, respectively; (c) DC varactor bias voltage $V_{DC}$, keeping the modulation frequency and modulation index to $f_m = 310$ MHz and $\Delta_m = 0.29$. In all cases, $\varphi_m$ has been set equal to 90 degrees.

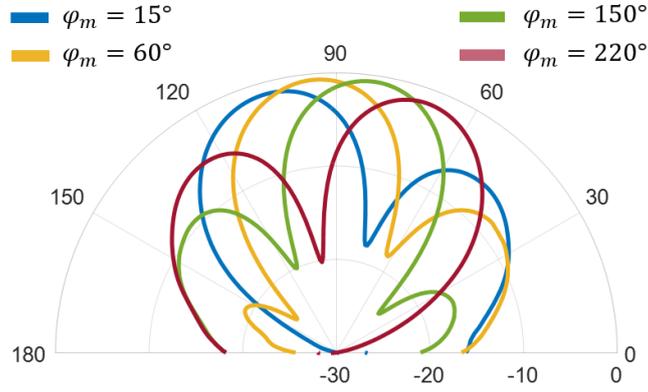

**Fig. 8. Beam scanning response of the nonreciprocal phased-array antenna described in Fig. 5 operated in transmission.** Results show the measured radiation diagram (in dB) in the E-plane at 2.4 GHz ($f_0 \rightarrow f_0 + f_m$) for various values of the phase difference ($\varphi_m$) between the modulation signals ($f_m = 310$ MHz) that control each element of the antenna array.

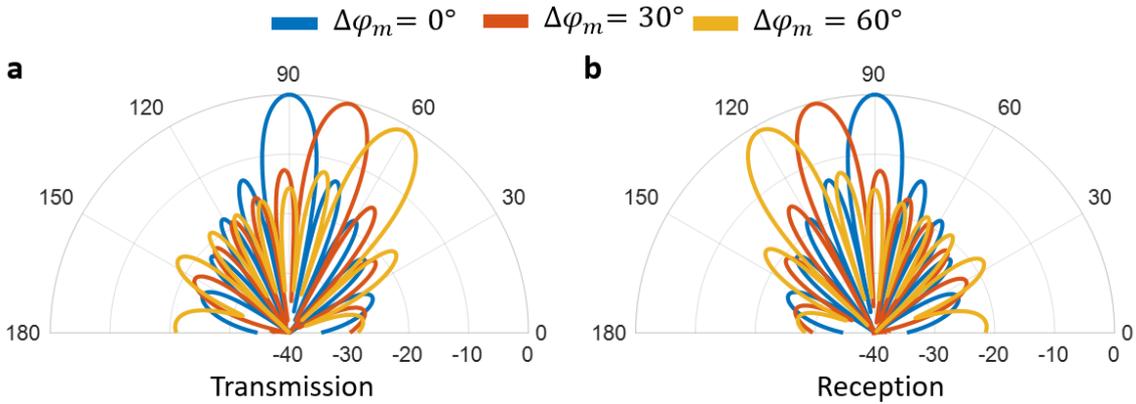

**Fig. 9. Numerical analysis of a nonreciprocal phased-array antenna composed of 1x8 elements and operating at 2.4 GHz.** The elements of the antenna are identical to the one described in Fig. 4 and the array follows the scheme of Fig. 1b-c, feeding all antennas with identical phase at $f_0$. E-plane radiation patterns (in dB) in (a) transmission and (b) reception operation are plotted versus the phase difference $\Delta\varphi_m$ imposed between adjacent antenna elements.

| $\varphi_m$ | Transmission ($f_0 \to f_0 + f_m$) | | Reception ($f_0 + f_m \to f_0$) | |
|---|---|---|---|---|
| | Left | Right | Left | Right |
| **0°** | 0° | −90° | 0° | −90° |
| **90°** | 0° | 0° | 0° | −180° |
| **180°** | 0° | 90° | 0° | −270° |
| **270°** | 0° | 180° | 0° | 0° |

**Table 1:** Phases imparted by the left and right patch elements of the antenna array described in Fig. 5 to the transmitted (left) and received (right) waves for several values of the phase difference between the low-frequency modulation signals that control each patch, $\varphi_m$. The initial phase difference between the left and right elements of −90 degrees is provided by the feeding network at $f_0$.

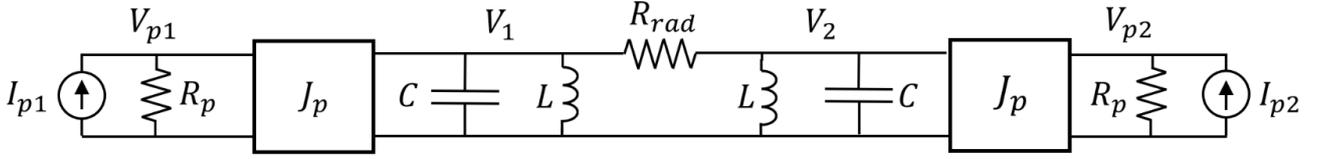

**Fig. B1. Equivalent circuit of a linear resonant antenna excited from two sides.** This circuit is similar to the one shown in Fig. 2a paper but incorporates an admittance inverter on each side to characterize the coupling between the port and the adjacent resonator. In addition, the different nodes are labelled to be consistent with the formulation detailed in Appendix B.

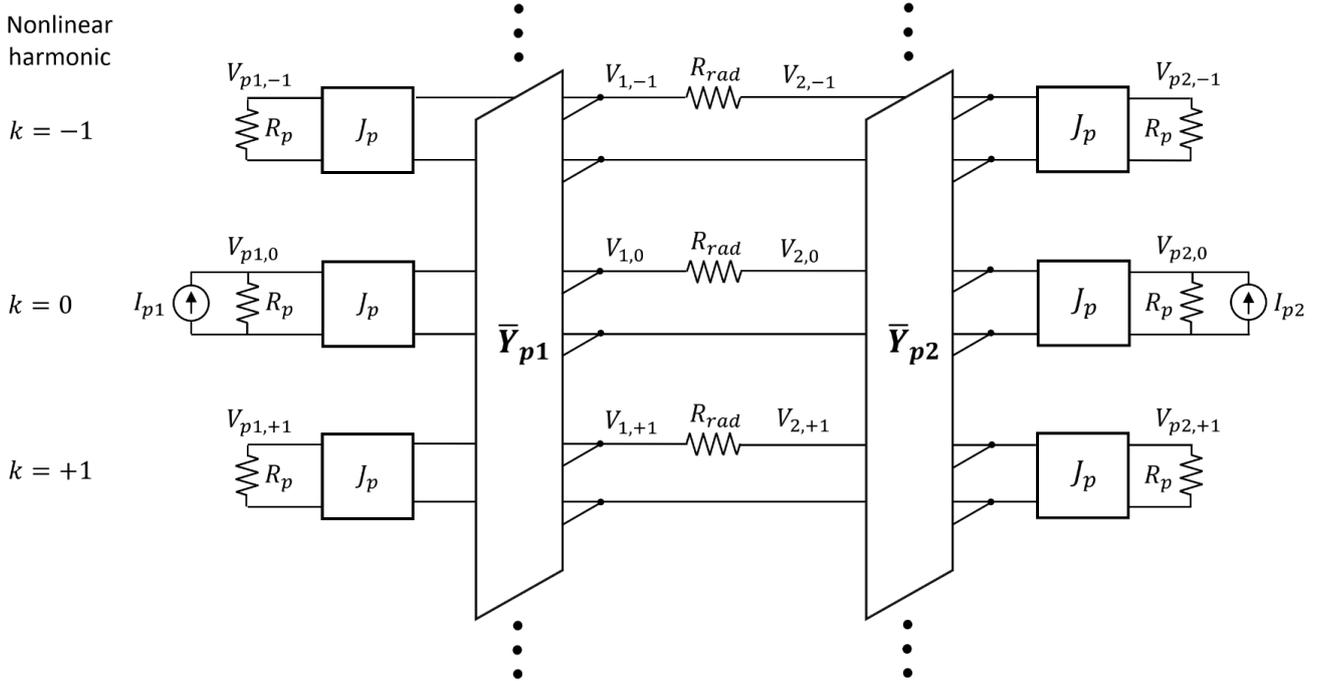

**Fig. B2. Equivalent circuit of the resonant antenna shown in Fig. B1 when their resonators are time-modulated as described in Fig. 2b.** The notation employed to denote the voltages on the node '$a$' follows the scheme $V_{a,k}$ where '$k$' is the order of a given nonlinear harmonic. The big boxes containing the matrixes $\overline{Y}_{p1}$ and $\overline{Y}_{p2}$ represent the admittance matrices of the time-modulated resonators that couples all nonlinear harmonics, as described in the Appendix B. Note that only the fundamental ($k = 0$) and first nonlinear harmonics ($k = \pm 1$) have been included in this schematic.

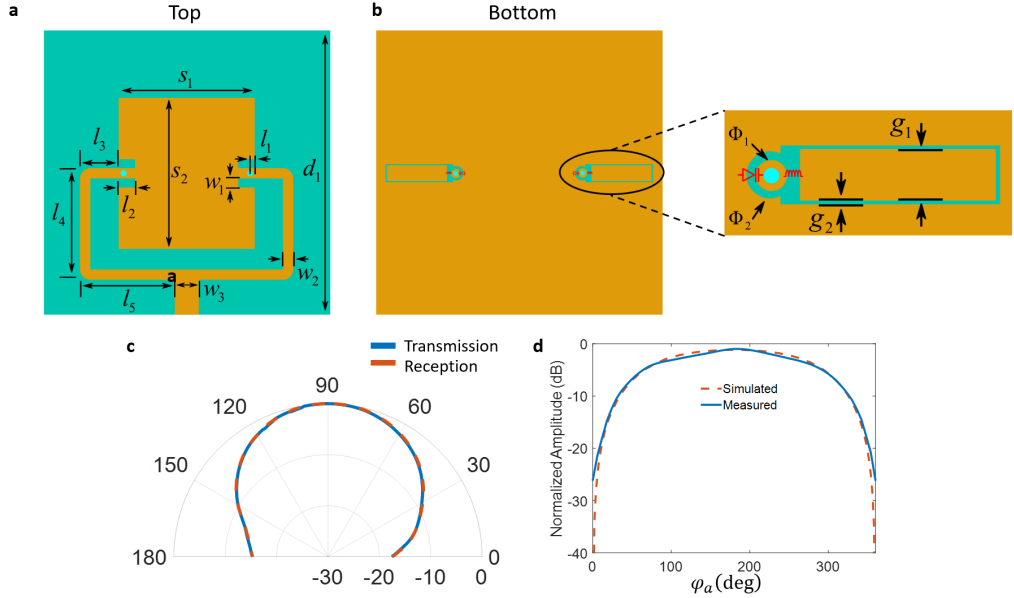

**Fig. C1. Details of the time-modulated patch antenna described in Fig. 4.** The antenna is fabricated using a dielectric Roger Corporation laminate RT/duroid 5880 with a thickness, permittivity, and tangent loss of $h = 1.575$ mm, $\varepsilon_r = 2.2$ and $\tan\delta = 0.0009$, respectively. (a) Top of the board. (b) Bottom of the board. The inset shows a close-up of the connection between the CPW (bottom) and the patch antenna (top) through a via-hole and the connection of the varactor (Skyworks SMV1233) employed to apply time-modulation and the inductor (TDK SIMID 33nH) that behaves as a choke. (c) Measured E-plane radiation pattern (in dB) in transmission and reception at 2.4 GHz. (d) Power transmitted by the antenna at the broadside direction measured in an anechoic chamber and normalized with respect to power of the standard patch antenna described in Fig. C3. Results are plotted versus the phase difference $\varphi_a$ imposed between the two modulation signals that control the varactors and show that maximum transmission power appears when this phase is 180°. Note that in Fig. 4 the phase $\varphi_a$ is set equal to 180 degrees. Other parameters are as follows: $l_1 = 2.5$ mm, $l_2 = 7$ mm, $l_3 = 11.2$ mm, $l_4 = 37.4$ mm, $l_5 = 32.3$ mm, $w_1 = 2.0$ mm, $w_2 = 1.4$ mm, $w_3 = 4.8$ mm, $s_1 = 47$ mm, $s_2 = 53$ mm, $d_1 = 100$ mm, $g_1 = 2.7$ mm, $g_2 = 0.2$ mm, $\Phi_1 = 1.0$ mm, and $\Phi_2 = 2.0$ mm.

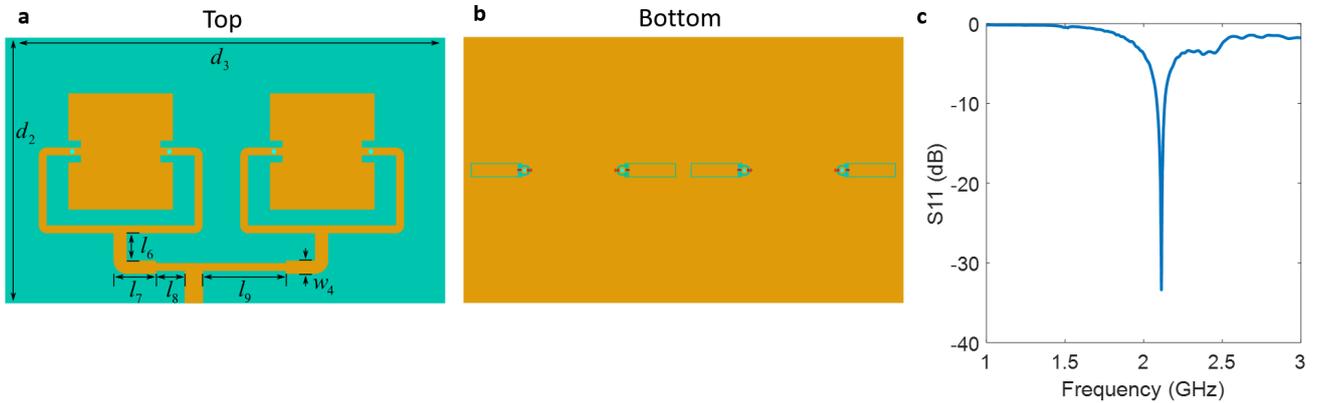

**Fig. C2. Details of the 2-element antenna array based on time-modulated patches described in Fig. 5.** The antenna is fabricated using a dielectric Roger Corporation laminate RT/duroid 5880 with a thickness, permittivity, and tangent loss of $h = 1.575$ mm, $\varepsilon_r = 2.2$ and $\tan\delta = 0.0009$, respectively. (a) Top of the board. (b) Bottom of the board. Each individual patch antenna, including CPWs lines, is identical to the one detailed in Fig. C1. (c) Measured scattering parameter S11 when the time-modulation described in Fig. 5 is applied. Other parameters are as follows: $l_6 = 6.9$ mm, $l_7 = 16.4$ mm, $l_8 = 17.6$ mm, $l_9 = 41.6$ mm, $w_4 = 2.8$ mm, $d_2 = 110$ mm and $d_2 = 194$ mm.

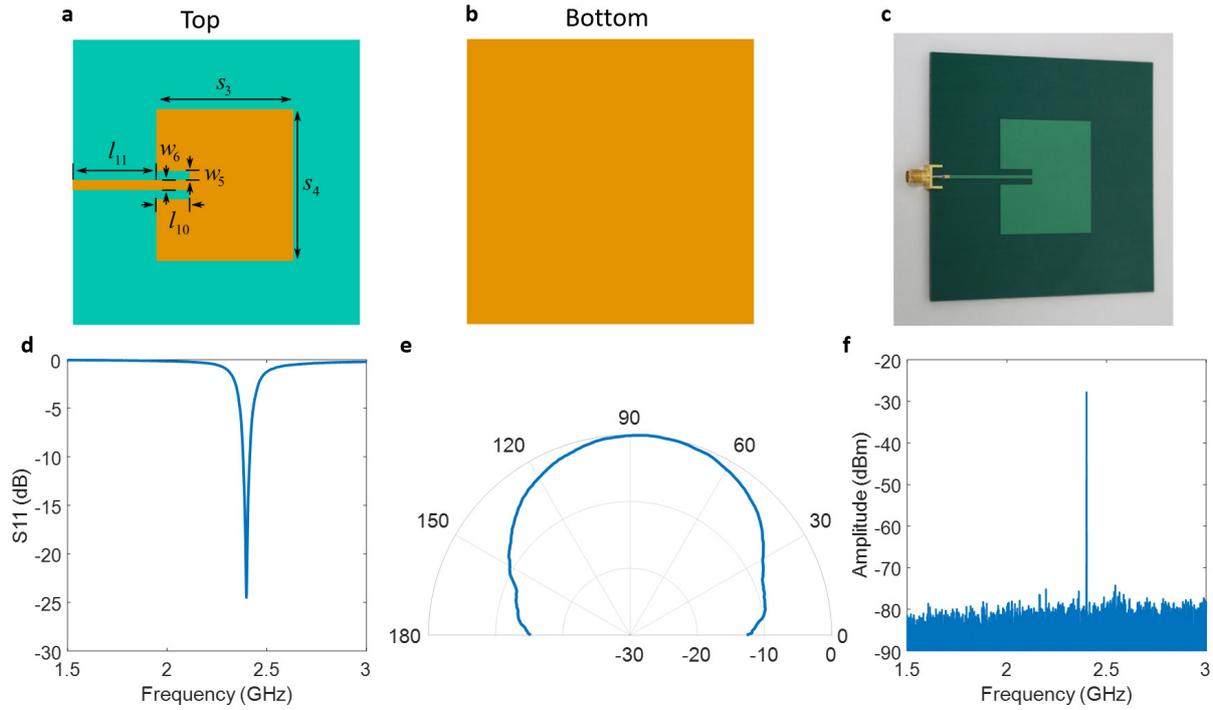

**Fig. C3. Details of the standard patch antenna employed as a reference in Fig. 4.** The antenna is fabricated using a dielectric Roger Corporation laminate RT/duroid 5880 with a thickness, permittivity, and tangent loss of $h = 1.575$ mm, $\varepsilon_r = 2.2$ and $\tan\delta = 0.0009$, respectively. (a) Top of the board. (b) Bottom of the board. (c) Picture of the fabricated prototype. (d) Measured scattering parameter S11. (e) Measured E-plane radiation pattern (in dB) at 2.4 GHz. (f) Spectrum of the power transmitted by the antenna at the broadside direction in an anechoic chamber. Other parameters are $l_{10} = 14$ mm, $l_{11} = 29.6$ mm, $w_5 = 2.5$ mm, $w_6 = 1.4$ mm, $s_3 = 40.8$ mm and $s_4 = 47.5$ mm.

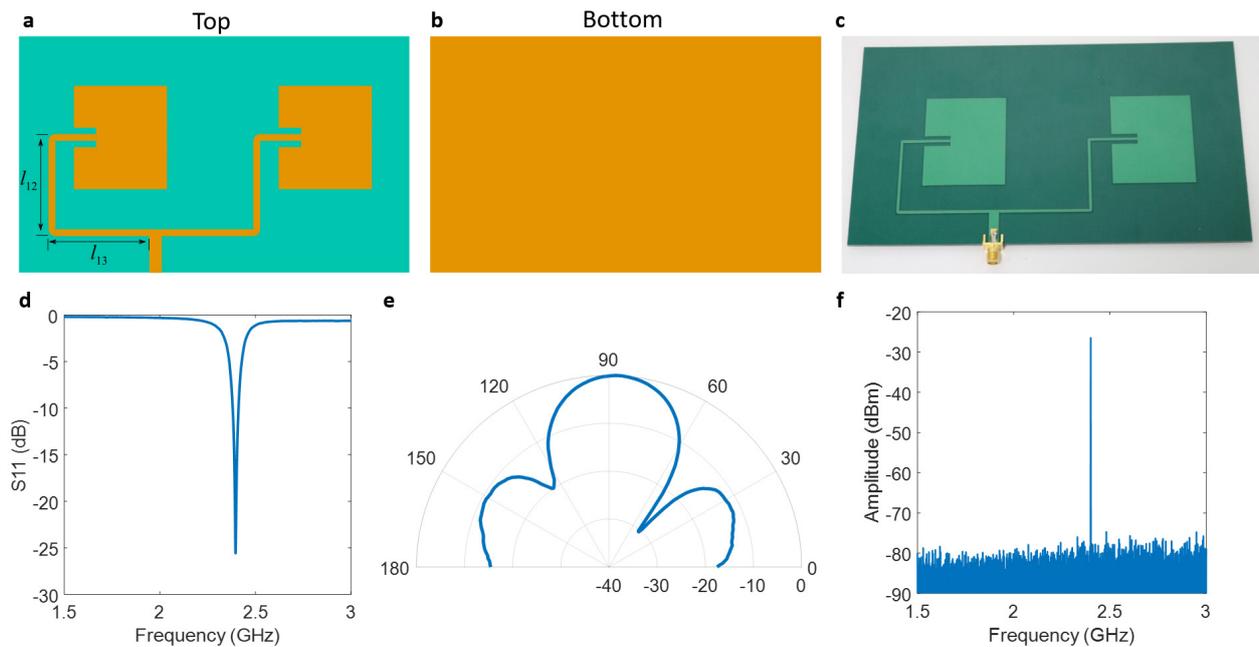

**Fig. C4. Details of the 2-element antenna array composed of patches employed as a reference in Fig. 5.** The antenna is fabricated using a dielectric Roger Corporation laminate RT/duroid 5880 with a thickness, permittivity, and tangent loss of $h = 1.575$ mm, $\varepsilon_r = 2.2$ and $\tan\delta = 0.0009$, respectively. (a) Top of the board. Each individual patch antenna is identical to the one detailed in Fig. C3. (b) Bottom of the board. (c) Picture of the fabricated prototype. (d) Measured scattering parameter S11. (e) Measured E-plane radiation pattern (in dB) at 2.4 GHz. (f) Spectrum of the power transmitted by the antenna at the broadside direction in an anechoic chamber. Other parameters are $l_{12} = 39.4$ mm and $l_{13} = 45.3$ mm.

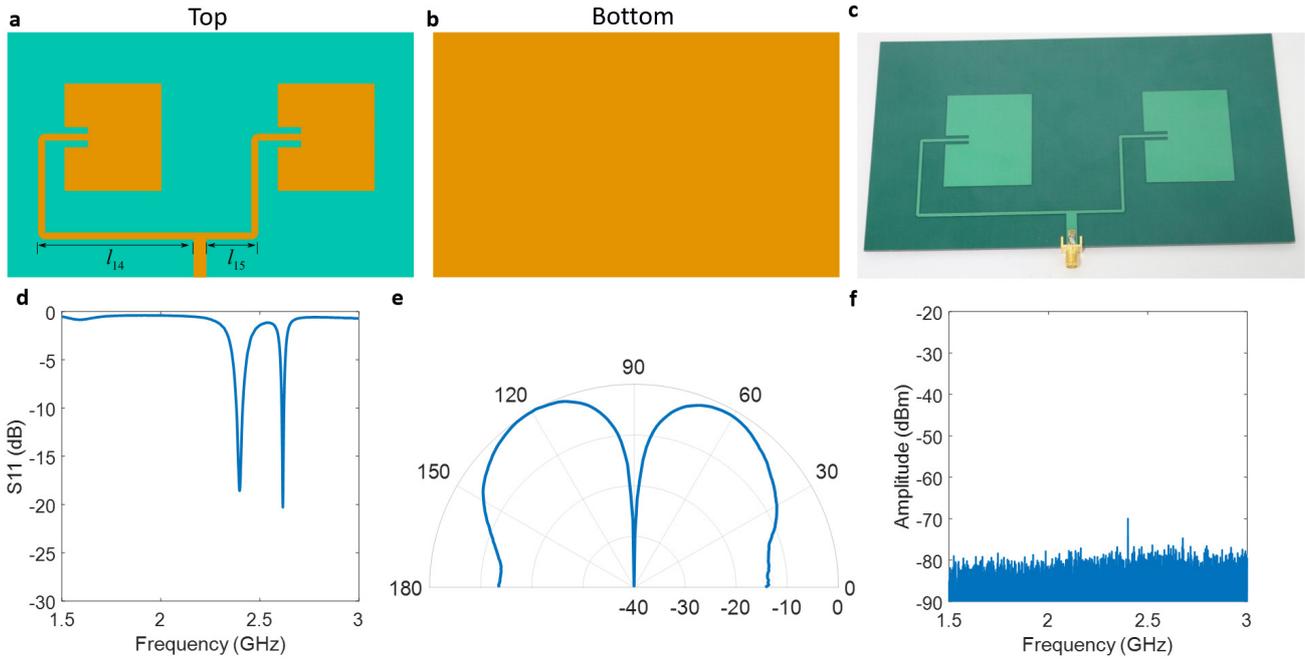

**Fig. C5. Details of a 2-element antenna array composed of patches that are fed with 180 phase difference.** The antenna is fabricated using a dielectric Roger Corporation laminate RT/duroid 5880 with a thickness, permittivity, and tangent loss of $h = 1.575$ mm, $\varepsilon_r = 2.2$ and $\tan\delta = 0.0009$, respectively. (a) Top of the board. Each individual patch antenna is identical to the one detailed in Fig. C3. (b) Bottom of the board. (c) Picture of the fabricated prototype. (d) Measured scattering parameter S11. (e) Measured E-plane radiation pattern (in dB) at 2.4 GHz. (f) Spectrum of the power transmitted by the antenna at the broadside direction in an anechoic chamber. Other parameters are $l_{14} = 68.7$ mm and $l_{15} = 21.9$ mm.